\documentclass[a4paper,referee]{raa}            % referee version: for submission
\usepackage[a4paper, top=2cm, bottom=3cm, left=2.5cm, right=2.5cm, columnsep=20pt]{geometry}

\usepackage{graphicx,times}
\usepackage{natbib}
\usepackage{amssymb,amsmath}
\bibpunct{(}{)}{;}{a}{}{,}

\usepackage[pagebackref=true]{hyperref}

\usepackage{multirow}
\usepackage{textcomp, gensymb}

\begin{document}

   \title{Long term X-ray spectral variations of the Seyfert-1 galaxy Mrk 279}

 \volnopage{ {\bf 20XX} Vol.\ {\bf X} No. {\bf XX}, 000--000}
   \setcounter{page}{1}

   \author{K. Akhila \inst{1}, 
    Ranjeev Misra \inst{2}, 
    Savithri H. Ezhikode \inst{3}, 
    K. Jeena \inst{1}
   }

\institute{Department of Physics, Providence Women's College, University of Calicut, Kerala 673009, India; {\it akhilasadanandan@gmail.com, jeenakarunakaran@gmail.com }\\
\and Inter-University Centre for Astronomy and Astrophysics (IUCAA), PB No.4, Ganeshkhind, Pune 411007, India\\
\and Department of Physics and Electronics, CHRIST (Deemed to be University), Hosur Main Road, Bangalore 560029, India\\
\vs \no
   {\small Received 20XX Month Day; accepted 20XX Month Day}
}

\abstract{We present the results from a long term X-ray analysis of Mrk 279 during the period 2018-2020. We use data from multiple missions - \textit{AstroSat}, \textit{NuSTAR} and \textit{XMM-Newton}, for the purpose. The X-ray spectrum can be modelled as a double Comptonisation along with the presence of neutral Fe K$\alpha$ line emission, at all epochs. We determined the source's X-ray flux and luminosity at these different epochs. We find significant variations in the source's flux state. We also investigated the variations in the source's spectral components during the observation period. We find that the photon index and hence the spectral shape follow the variations only over longer time periods. We probe the correlations between fluxes of different bands and their photon indices, and found no significant correlations between the parameters.
\keywords{galaxies:Seyfert – galaxies:individual:Mrk 279 - X-rays:galaxies 
}
}
   \authorrunning{K. Akhila et al. }            %author_head in even pages
   \titlerunning{X-ray spectral variations of Mrk 279}  % title_head in odd pages
   \maketitle
%________________________________________________ sections below
% 
%%%%%%%%%%%%%%%%%%%%%%%%%%%%%%%%%%%%%%%%%%%%%%%%%%%%%%%%%%
\section{Introduction}           %% first-level sections will be auto-capitalized
\label{sect:intro}

Seyfert-1 galaxies are the low-luminosity sub-class of active galaxies characterised by the presence of both broad and narrow emission lines in their spectra \citep[E.g.,][]{schmidt1983quasar, netzer2015revisiting}. The typical X-ray spectrum of a narrow line Seyfert 1 (NLS1) galaxy is a power law continuum with several features like the presence of broad and narrow emission lines and a smooth excess emission component at soft energies \citep{osterbrock1985spectra, arnaud1985exosat, boller1995soft, fabian2000broad, rani2019study, ezhikode2021astrosat, mochizuki2023origin}. The current understanding of the AGN structure tells us that these features arise from regions of varying environmental conditions, of significantly different temperatures and densities, around the central supermassive black hole.

All classes of AGNs are known to have variations in their X-ray fluxes. These variations occur over a wide range of timescales and amplitudes. AGN X-ray emission have been observed to exhibit variations of over a few $\sim1000 s$ upto years with amplitude variations upto an order of magnitude \cite[E.g.,][]{turner1999x}. Studies have shown that majority of NLS1 galaxies soften as they become brighter and that the variability is stronger in softer energies \citep{markowitz2003long, markowitz2004expanded}.

Markarian 279 (Mrk 279) is a nearby NLS1 galaxy at a redshift of z = 0.0305 \citep{scott2009variable} harboring a supermassive black hole of mass M = $3.49 \times 10^{7} M_\odot$ at its centre \citep{peterson2004central}. It has been observed extensively in the X-ray energies by various missions. Early observations were made by \textit{HEAO 1} and Japan's \textit{ASCA} mission \citep{dower1980seyfert, weaver1995confirmation, weaver2001variable}. \textit{Chandra}, \textit{HST-STIS} and \textit{FUSE} observed the source simultaneously in May 2002 \citep{scott2004intrinsic, kaastra2004x, arav2005x} and again in 2003 \citep{gabel2005x, arav2007chemical, costantini2007x}. Mrk 279 has also been observed multiple times by \textit{Chandra} and \textit{XMM-Newton}, independently \citep{yaqoob2004cores, jiang2019high, igo2020searching, ursini2020estimating}. These observations have revealed that the X-ray flux of the source show significant variations over time \citep{scott2004intrinsic, costantini2010xmm, ebrero2010xmm}.

Long term X-ray observations of Mrk 279 from 1979 to 1994 reveal a continuum flux variability going upto a factor of five, ranging between $1-5 \times 10^{-11}$ erg $cm^{-2}$ $s^{-1}$ \citep{weaver2001variable}. \cite{weaver2001variable} analysed the \textit{ASCA} observation from 1994 using a model consisting of two power laws and a narrow Gaussian. This showed that the 2-10 keV flux increased by $20 \%$ over the period of a few hours. In 2002, analysis of the \textit{Chandra} observation using a continuum model consisting of two power laws modified by Galactic absorption found a low 2-10 keV flux of $1.2 \times 10^{-11}$ erg $cm^{-2}$ $s^{-1}$ \citep{scott2004intrinsic}. Such low flux levels were previously observed in 1979 and 1991 \citep{weaver2001variable}. The flux had dropped by a factor of two from a previous \textit{XMM-Newton} observation in the same year. \cite{scott2004intrinsic} also reported the UV continuum flux to have decreased by a factor of $\sim 7.5$ from 1999 to 2002. \textit{XMM-Newton} has further observations of the source spread over multiple orbits during November, 2005. \cite{costantini2010xmm} found that at least three components are needed to give an acceptable description of the continuum spectrum. They fitted the epic-pn data using a broken power law and a modified black body component. They calculated the 2-10 keV flux to be $\sim 2.5 \times 10^{-11}$ erg $cm^{-2}$ $s^{-1}$. The \textit{Swift}-BAT survey yielded a broad band flux of $\sim 3.8 \times 10^{-11}$ erg $cm^{-2}$ $s^{-1}$ in the 15 - 150 keV X-ray band \citep{cusumano2010palermo}.

Here, we use data from multiple missions to analyse the X-ray emissions from the source for a three year period extending from early 2018 to late 2020, so as to study its flux state and the variations that happen therein. This paper is structured as follows. The details of the observations used are given in Section \ref{sect:obs}. Section \ref{sect:datared} discusses the processing techniques adopted for each data set. Spectral analysis is given in Section \ref{sect:spec} and we analyse the flux variations of the source in Section \ref{sect:flux}. Finally the results are discussed in Section \ref{sect:res}.

%%%%%%%%%%%%%%%%%%%%%%%%%%%%%%%%%%%%%%%%%%%%%%%%%%%%%%%%%%%%%%%%
\section{Observations}
\label{sect:obs}
\textit{AstroSat}, India's first multi wavelength space observatory, enables simultaneous observations in the broad X-ray and UV bands. In this analysis, we use the first and so far, the only \textit{AstroSat} observation of the source made in 2018. We use the data from the Soft X-ray Telescope and Large Area X-ray Proportional Counter instruments of \textit{AstroSat} (\textit{AstroSat} : \citealt{agrawal2006broad}, SXT : \citealt{singh2017soft}, LAXPC : \citealt{yadav2016large, antia2017calibration, agrawal2017large, misra2017astrosat}). The NUclear Spectroscopic Telescope Array (\textit{NuSTAR}) has made observations of the source in the years 2019 and 2020. We chose the observation from 2019 and three orbits with enough exposure time in 2020 by the Focal Plane Modules (FPMA and FPMB) of \textit{NuSTAR} \citep{harrison2010nuclear, harrison2013nuclear}. Even though the \textit{NuSTAR} observation from 2020 has been previously studied for multiple purposes \citep{akylas2021distribution, 2022A&A...666A.127A, kang2022x, pal2023properties}, our aim here is to analyse it for the variations in its X-ray flux state. We also use data from the epic-pn camera of the X-ray Multi-mirror Mission (\textit{XMM-Newton} :  \citealt{jansen2001xmm, struder2001european}) for this analysis. \textit{XMM-Newton} has observed the source several times in the period 2002-2020; the earlier observations have been widely analysed, which is already discussed in Section \ref{sect:intro}. We selected the latest \textit{XMM-Newton} observation with a $30.5$ ks long exposure from 2020, the analysis of which has not been reported before. The details of all chosen data sets are summarised in Table \ref{tab:obslog}.

\begin{table}
\centering
\caption{Observation log detailing the date of observation, satellite, mission id and exposure time.}
\label{tab:obslog}
\begin{tabular}{cccc}
\hline\hline
Mission & \begin{tabular}[c]{@{}c@{}}Observation\\ Id\end{tabular} & \begin{tabular}[c]{@{}c@{}}Date of\\ Observation\end{tabular} & \begin{tabular}[c]{@{}c@{}}Exposure\\ Time (ks)\end{tabular} \\
\hline
\textit{\begin{tabular}[c]{@{}c@{}}AstroSat\\ SXT, LAXPC\end{tabular}} & 9000001886 & 06-Feb-2018 & 39.94,101 \\
\multirow{4}{*}{\textit{\begin{tabular}[c]{@{}c@{}}NuSTAR\\ FPMA, FPMB\end{tabular}}} & 60160562002 & 29-Oct-2019 & 27.27 \\
 & 60601011002 & 03-Aug-2020 & 62.02 \\
 & 60601011004 & 05-Aug-2020 & 200.63 \\
 & 60601011006 & 11-Aug-2020 & 52.80 \\
\textit{\begin{tabular}[c]{@{}c@{}}XMM-Newton\\ epic-pn\end{tabular}} & 0872391301 & 20-Dec-2020 & 20.02 \\
\hline
\end{tabular}
\end{table}
%%%%%%%%%%%%%%%%%%%%%%%%%%%%%%%%%%%%%%%%%%%%%%%%
\section{Data Reduction}
\label{sect:datared}

\subsection{AstroSat}
The Soft X-ray Telescope (SXT) and Large Area X-ray Proportional Counters (LAXPCs) instruments of \textit{AstroSat} observe X-ray sources in the $0.3$ - $8.0$ keV and $3.0$ - $80.0$ keV energy bands, respectively. SXT observations were made in the Photon Counting (PC) mode of the telescope. Level-1 data from 20 orbits of \textit{AstroSat}-SXT were processed using the \textsc{sxtpipeline 1.4b} software and level-2 data was extracted. Data corresponding to all 20 orbits were then merged into a single cleaned event file. For this the event merger tool, \textsc{sxtevtmerger}, was used. The \textsc{xselect} task of HeaSoft was used for filtering the data. A circular region of radius $15'$ was chosen around the source and the spectrum was extracted. Blank sky background was used for the background spectrum. Response files (ARF and RMF) for the PC mode of the telescope, provided by SXT-POC team, were used during spectral fitting.

LAXPC is a cluster of three identical proportional counters performing X-ray observations in a fairly broad energy band. We use data from LAXPC20 alone for our analysis as it has more stability in its response. The data was processed using \textsc{laxpcsoft}. The software's tools were used to extract level-2 data, from which a fits event file was created. A filter file containing the good time intervals was also generated. The source and blank sky background spectra and the response files were extracted using the LAXPCSoftware.

\subsection{NuSTAR}
\textit{NuSTAR} is an X-ray observatory operating in the $3$-$79$ keV energy range. We use data from both the telescopes, FPMA and FPMB, for our analysis. NuSTAR Data Analysis Software \textsc{nustardas\_v2.1.1} package was used for processing the data. The \textsc{nupipeline} was run using the latest calibration (CALDB) files. Output files were created and these were calibrated and cleaned using the created gti files. A circular region of size $38"$ was selected around the source and another region of similar size was selected for background extraction. Then the \textsc{nuproducts} task was run to extract spectral files and lightcurves. Science products were extracted for data from both the telescopes.

\subsection{XMM-Newton}
\textit{XMM-Newton} is an X-ray observatory which make observations in the $0.1$ - $12$ keV energy band. We use the data from the epic-pn camera taken in the imaging mode for our analysis. The Science Analysis Software \textsc{sas-20.0.0} was used for processing the data. We first created a raw event file using the \textsc{epchain} task and filtered the event file to remove flaring backgrounds. Then a gti file was created using \textsc{tabgtigen}. Only those data satisfying the conditions \textsc{pattern}$\leq$4 and \textsc{flag}==0 were selected. \textsc{SAS} task \textsc{epatplot} was used to verify that the observation was free from any pile-up effect. We selected a circular region, centered around the source, with a radius of $40"$ and the spectrum was extracted. Similarly, background spectra were extracted from regions of the same size nearby, but excluding the source. Response files were generated using the tasks \textsc{arfgen} and \textsc{rmfgen}.

%%%%%%%%%%%%%%%%%%%%%%%%%%%%%%%%%%%%%%%%%%%%
\section{Spectral Analysis}
\label{sect:spec}
Spectral fitting was done using the $\chi^2$ statistic. The spectral files were grouped so as to have a minimum of 30 counts in each bin. As advised by the LAXPC team, its spectrum was grouped at a 5\% level, giving three energy bins per resolution. We use HeaSoft's spectral fitting package, \textsc{xspec}-12.12. \citep{arnaud1996astronomical}. The errors on the model parameters, obtained from \textsc{xspec}, give their 90\% confidence interval. We fit the spectrum using two Comptonization components and a Gaussian component for the iron K-$\alpha$ emission line. Such a two corona model is widely used in literature to describe the observed X-ray spectrum of several NLS1 galaxies \citep{magdziarz1998spectral, czerny2003universal, done2012intrinsic, petrucci2018testing, garcia2019implications}. In this, the primary hard X-ray continuum is modelled to arise from the hot corona, explained as a powerlaw emission due to a Comptonisation component. In addition, there is another thermal Comptonisation of seed photons from the far UV end of accretion disk giving rise to the soft excess component. This is produced from a warm, optically thick layer of gas above the surface of the accretion disk and is distinct from the hot corona.

We use the \textit{diskbb} model of \textsc{xspec} for the black body continuum from the accretion disk and this is convolved by \textit{thComp} model of \textsc{xspec}. \textit{thComp}, a convolution model which can be used for any seed photon distribution, is a replacement for the older \textit{nthComp} model and agrees better with the Monte Carlo results for thermal Comptonisation \citep{zdziarski2020spectral}. A fraction of the photons thus Comptonised are further upscattered by another Comptonisation medium represented by the \textsc{xspec} model \textit{simpl} \citep{steiner2009simple}. \textit{simpl} is again a simple convolution model that employs just two parameters to model Compton scattering, that makes it appropriate to be used in situations when the temperature of the corona is high and cannot be constrained by the data being fit. The Galactic absorption is modelled using \textit{TBabs} model of \textsc{xspec} with the absorption column density fixed to the value, $N_H$ = $1.78\times10^{20}$ $cm^{-2}$ \citep{williams2006chandra}. We added the \textsc{xspec} model \textit{ztbabs} to model the intrinsic absorption of the source. We found that this does not improve the fit and using  more complicated models did not seem warranted and hence this component was not used in further analysis. Redshift of the source is fixed at z = $0.0304$. Assuming a typical inner radius of $\sim10R_g$ for the accretion disc of an NLS1 galaxy with a black hole mass $3.49 \times 10^{7} M_\odot$ \citep{peterson2004central}, we fix the normalisation of the \textit{diskbb} component to the value $5.90\times10^{9}$. Here we consider the source to be at a distance of $124.1$ Mpc \citep{pogge2002hubble} with an inclination angle of $30$\degree \citep{weaver2001variable}. Covering fraction of \textit{thComp} is kept at unity so that all the seed photons from the disk undergo Compton upscattering. 

\begin{figure}
    \centering
    \includegraphics[width=0.6\linewidth]{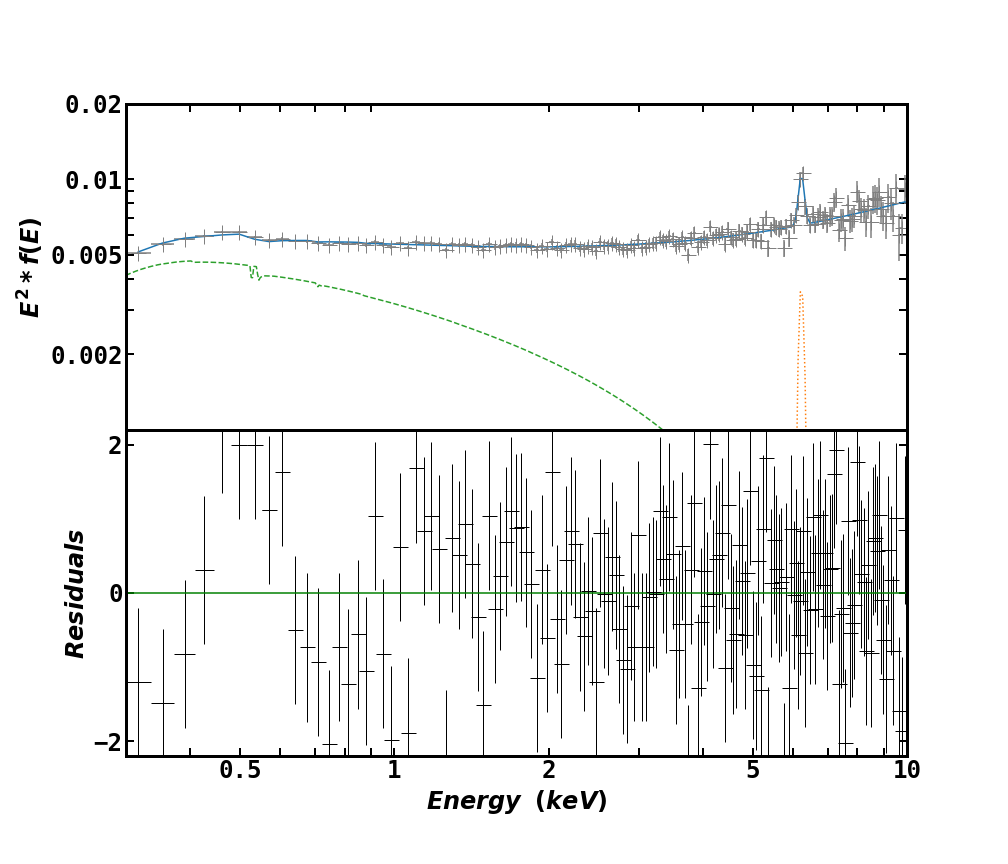}
    \caption{\textit{XMM-Newton} spectrum for the energy range 0.3 - 10.0 keV fit using theoretical model. Bottom panel shows the residuals of fit.}
    \label{fig:xmm_spec}
\end{figure}

We fit the 0.3-10.0 keV \textit{XMM-Newton} spectrum observed in December 2020 using this model. The spectrum is plotted in Figure \ref{fig:xmm_spec} and the best fit values of the parameters are listed in Table \ref{tab:xmm_spec}.

In order to probe the variations in corona over the years, we first fix the accretion disk as modelled by \textit{XMM-Newton} data. Accordingly the inner disk temperature ($T_{in}$) is kept fixed at $4.8\times10^{-3}$ keV. We also keep the optical depth of the warm corona and Gaussian line energy fixed at $15.02$ and $6.41$ keV, respectively. We then use this model to fit \textit{AstroSat}'s combined SXT-LAXPC spectrum and the \textit{NuSTAR} spectra. \textit{AstroSat} SXT and LAXPC spectra are in the energy ranges 0.3-6.0 keV and 4.0-20.0 keV, respectively while \textit{NuSTAR} covers a broad energy range of 3.0-79.0 keV. For the \textit{AstroSat} spectrum, we added the \textit{constant} model of \textsc{xspec} to accommodate the cross normalisation between SXT and LAXPC20.

\begin{figure}
    \centering
    \includegraphics[width=0.6\linewidth]{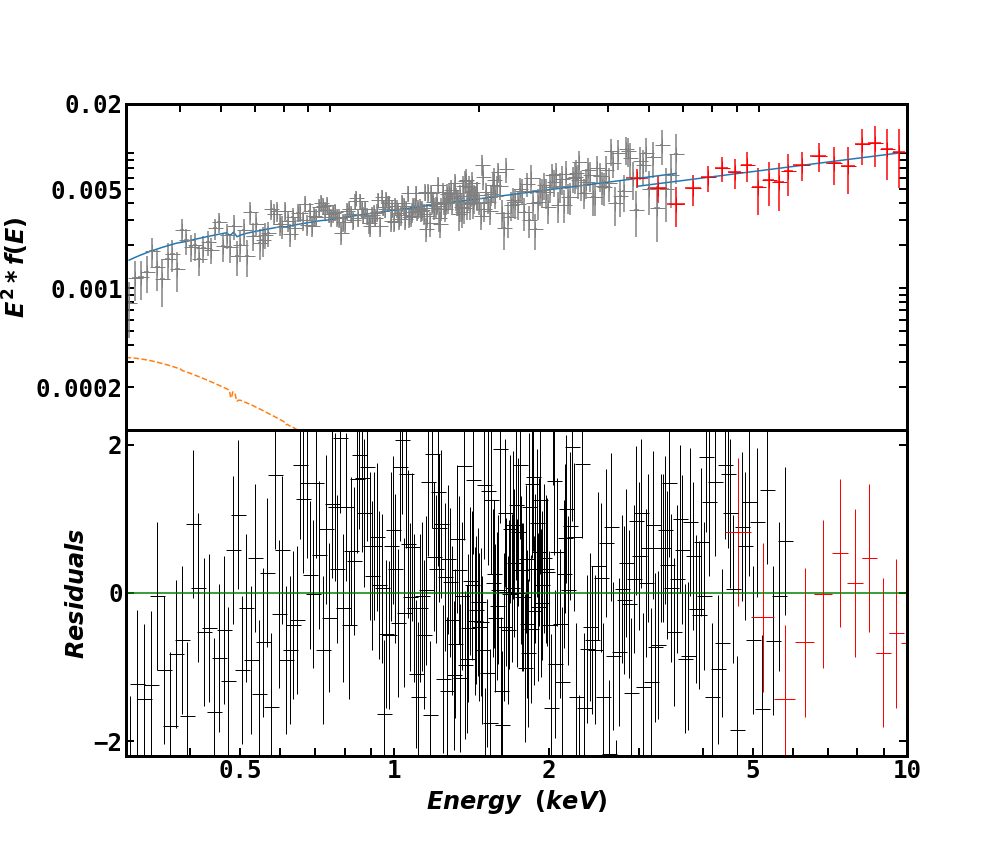}
    \caption{Combined SXT - LAXPC spectrum fit using theoretical model. Bottom panel shows the residuals of fit.}
    \label{fig:astro_spec}
\end{figure}

\begin{figure*}
    \centering
    \begin{subfigure}{0.48\textwidth}
        \includegraphics[width=\linewidth]{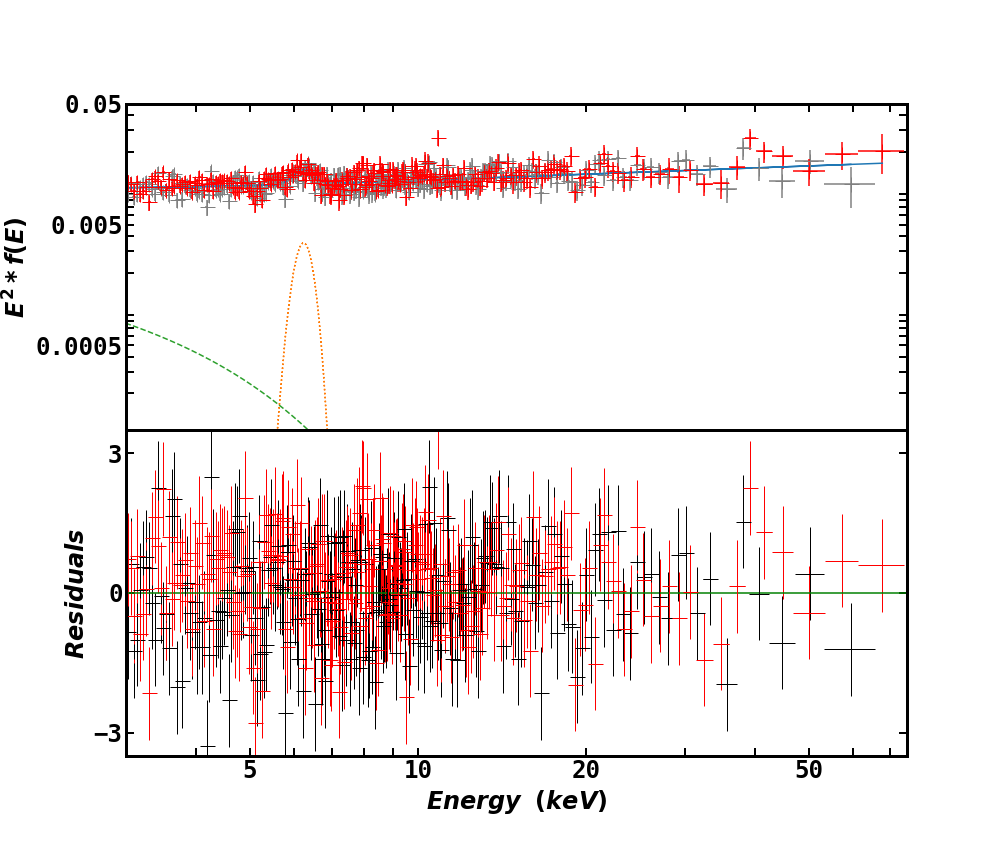}
        \caption{}
        \label{fig:nu1}
    \end{subfigure}
    \begin{subfigure}{0.48\textwidth}
        \includegraphics[width=\linewidth]{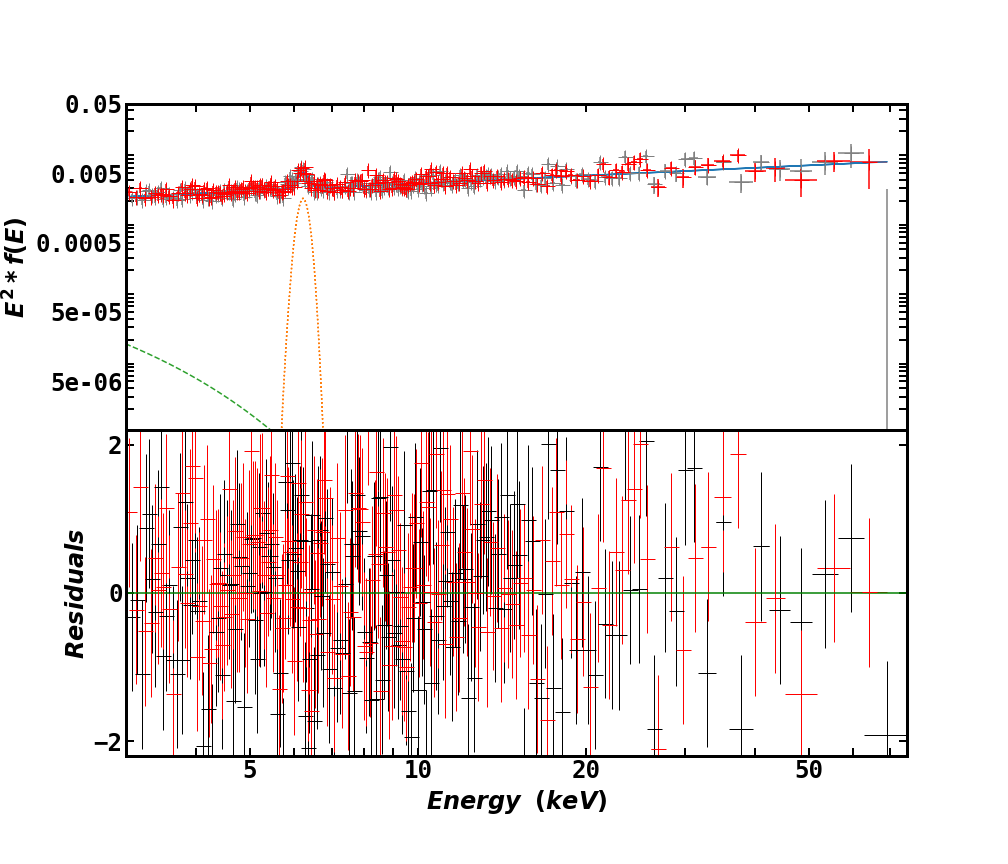}
        \caption{}
        \label{fig:nu2}
    \end{subfigure}
    \begin{subfigure}{0.48\textwidth}
        \includegraphics[width=\linewidth]{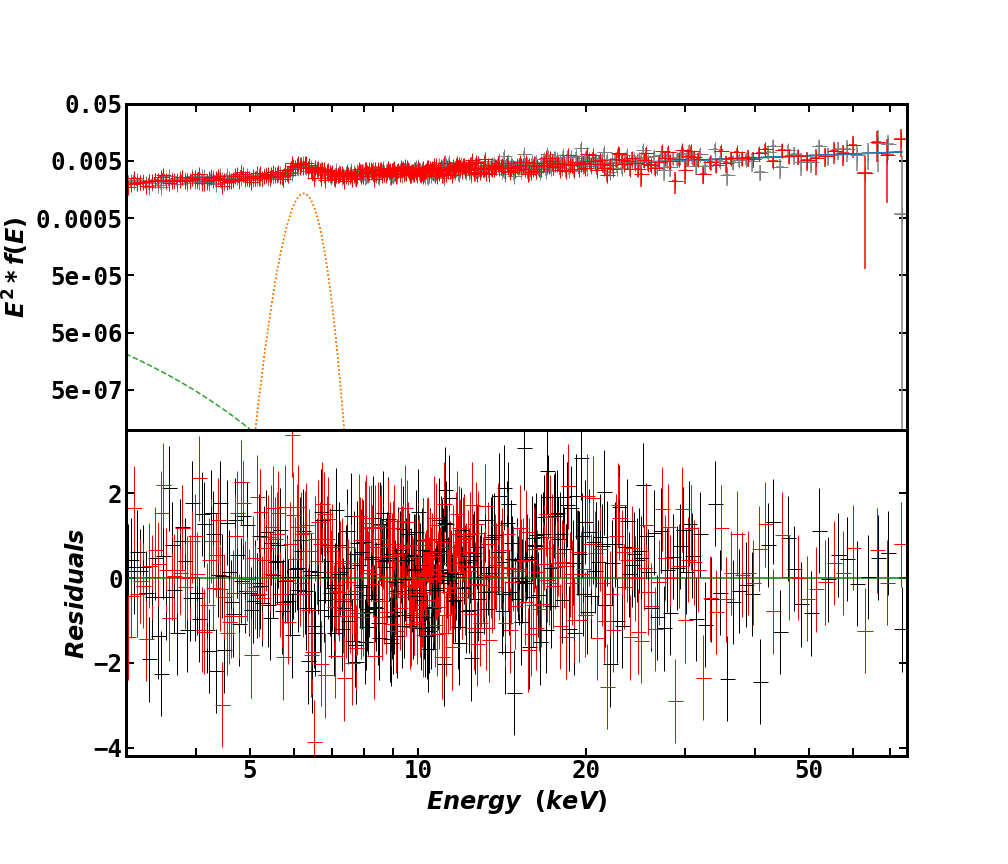}
        \caption{}
        \label{fig:nu3}
    \end{subfigure}
    \begin{subfigure}{0.48\textwidth}
        \includegraphics[width=\linewidth]{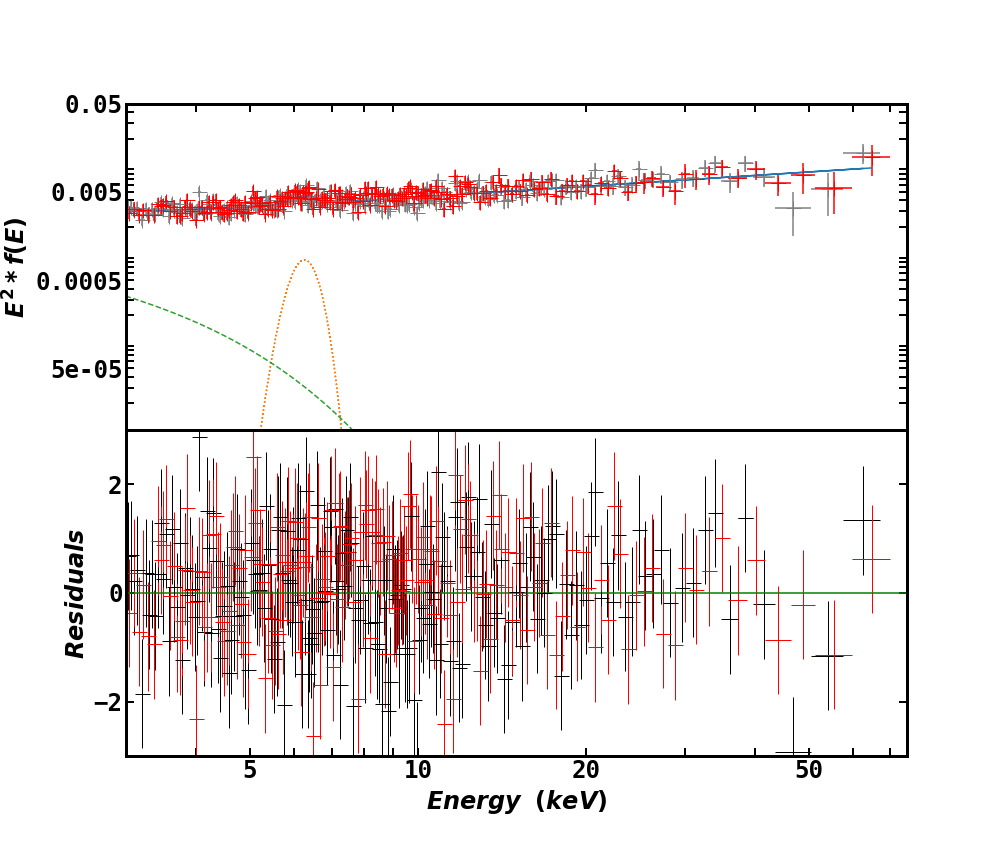}
        \caption{}
        \label{fig:nu4}
    \end{subfigure}    
    \caption{The \textit{NuSTAR} 3.0 - 79.0 keV spectra fit using the theoretical model. Top panels shows the unfolded spectrum along with the model while bottom panel plots the residuals of fit. Starting from the top left corner, the different figures correspond to the (a) 60160562002, (b) 60601011002, (c) 60601011004 and (d) 60601011006 \textit{NuSTAR} observations, respectively.}
    \label{fig:nu_spec}
\end{figure*}

The free parameters of fit for \textit{AstroSat} and \textit{NuSTAR} data sets are listed in Table \ref{tab:all_spec}. We find that the photon index for the hot corona ($\Gamma_{simpl}$) undergoes considerable changes during the period of observation. \textit{AstroSat} data gives a slightly higher value ($\sim 1.62$) compared to \textit{XMM-Newton}. Analysing the \textit{NuSTAR} data, we notice that the index increases to $\sim 1.8$ for the observation from 2019 but then drops to $\sim 1.6$ for the three observations in August 2020. We find that the scattering fraction also follows a similar pattern. We were also able to constrain the warm coronal temperature ($kT_{e}$) with an upper bound for all data sets. The \textit{AstroSat} spectrum is not able to properly resolve the Gaussian line, nevertheless we obtained an upper constraint on its strength. The combined \textit{AstroSat} spectrum is plotted in Figure \ref{fig:astro_spec} while Figure \ref{fig:nu_spec} plots all the \textit{NuSTAR} spectra.

\begin{table}
    \centering
    \caption{\textit{XMM-Newton} spectral fitting parameters for the broadband 0.3 - 10.0 keV spectrum.}
    \label{tab:xmm_spec}
    \begin{tabular}{c c c}
    \hline
    \hline
   Model & Parameter & \textit{XMM-Newton}\\
   Component & Value & 0872391301\\
    \hline
    simpl   & $\Gamma_{simpl}$ & $1.48_{-0.21}^{+0.16}$\\
            & $f_{Scat}$($\times10^{-2}$) & $0.72_{-0.39}^{+0.61}$\\
    thComp & $\tau$ & $15.02_{-10.66}^{+2.38}$\\
            & $kT_e$ (keV) & $0.98_{-0.34}^{+2.09}$\\
    diskbb  & $T_{in}$($\times10^{-2}$ keV) & $0.48_{-0.03}^{+0.02}$\\
    zgauss  & $E_l$ (keV) & $6.41_{-0.04}^{+0.04}$\\
            & $\sigma$ (keV) & $<0.14$\\
            & $N_{gauss}(\times10^{-5})$ & $1.98_{-0.53}^{+0.57}$\\
    
    $\chi^2$/dof& & $158.66/159$\\
    \hline
    \end{tabular}
\end{table}

\begin{table*}
    \centering
    \caption{Spectral parameter values for the \textit{AstroSat} and \textit{NuSTAR} observations. All the other parameters are kept fixed at the values obtained from the best fit model of \textit{XMM-Newton}.}
    \label{tab:all_spec}
    \begin{tabular}{c c c c c c c}
    \hline
    \hline
   Model & Parameter & \textit{AstroSat} & \multicolumn{4}{c}{\textit{NuSTAR}}\\
   Component & Value & 9000001886 & 60160562002 & 60601011002 & 60601011004 & 60601011006\\
    \hline
    simpl   & $\Gamma_{simpl}$ & $1.62_{-0.04}^{+0.04}$ & $1.83_{-0.04}^{+0.04}$  & $1.63_{-0.04}^{+0.02}$ & $1.62_{-0.01}^{+0.01}$ & $1.59_{-0.05}^{+0.04}$ \\
            & $f_{Scat}$($\times10^{-2}$) & $1.46_{-0.30}^{+0.37}$ & $5.83_{-1.39}^{+3.69}$  & $0.76_{-0.19}^{+0.06}$ & $0.70_{-0.15}^{+0.39}$ & $0.70_{-0.18}^{+0.26}$ \\
    thcomp  & $kT_{e}$ (keV) & $<0.52$ &  $<1.02$ & $<0.81$ & $<0.77$ & $0.83_{-0.25}^{+0.07}$ \\
    zgauss  & $\sigma$ (keV) &  $0.08$(frozen) & $0.24_{-0.1}^{+0.1}$  & $<0.23$ & $0.27_{-0.06}^{+0.07}$ & $0.36_{-0.21}^{+0.39}$ \\
            & $N_{gauss}$ ($\times10^{-5}$) &  $<3.20$ &  $5.62_{-1.42}^{+1.54}$ & $1.95_{-0.45}^{+0.49}$ & $2.33_{-0.29}^{+0.38}$ & $1.96_{-0.79}^{+1.28}$ \\
    $\chi^2$/dof& & $271.96/270$ &  $502.32/509$ & $373.98/396$ & $762.99/740$ & $379.15/403$ \\
    \hline
    \end{tabular}
\end{table*}

%%%%%%%%%%%%%%%%%%%%%%%%%%%%%%%%%%%%%%%%%%%%%%%%%%
\section{Variations in coronal X-ray emission}
\label{sect:flux}
From the spectra, we calculated the source's flux and hence luminosity in the $0.001$ - $100.0$ keV energy range at all the different epochs. This is tabulated in Table \ref{tab:lum}. We find that the source goes through large variations in its continuum flux over the three year time scale. The source flux increases by more than $50\%$ from February 2018 to October 2019. It then falls into a state of very low flux, with the flux values dropping by about $2$ times. \textit{XMM-Newton} observations from late 2020 shows that the source has once again regained the flux, returning to its brighter state. Despite the changes in its flux states throughout the observation period, the source is constantly in a brighter state, with the flux values being several times higher, compared to the earlier observations mentioned in Section \ref{sect:intro}. Recent insights into the cross calibration issue between \textit{NuSTAR} and \textit{XMM-Newton} epic suggests an empirical correction be added to the effective area \citep{furst2022xmm, kang2023joint}. We note that the epic fluxes being lower by about 20\%, as mentioned by the \textit{XMM-Newton} team, would not have a significant effect on the calculated flux for the current observation.

We then use the \textit{cflux} model of \textsc{xspec} to individual model components. Keeping the value of fractional scattering ($f_{Scat}$) fixed, we apply \textit{cflux} to the entire Comptonisation term, \textit{simpl*thComp*diskbb} and then to \textit{thComp*diskbb} alone. The former term gives the value of the flux we observe as being due to the two coronal components, the net continuum flux. It includes the flux of photons that underwent scattering by both coronae plus the flux from the remaining photons which were scattered by the warm corona alone. We call this flux the net continuum flux (NCM), since it does not include the line emission. The latter term gives the total flux upscattered from the warm corona. It would have been observed as being the result of upscattering by the warm corona, had the hot corona not been present and is indicative of the properties of the warm coronal component (WCM : \textit{cflux*thcomp*diskbb}). Both the terms include the flux from the disk component. This is estimated by applying \textit{cflux} to \textit{diskbb} alone. The flux obtained from the accretion disk is $6.46\times10^{-11}$ erg $cm^{-2}$ $s^{-1}$. Applying \textit{cflux} to the Gaussian component (\textit{zgauss}), with the normalisation ($N_{Gauss}$) fixed, gives the emission line flux. This Gaussian line corresponding to iron K$\alpha$ emission contribute very little to the overall flux, its strength being roughly around $\sim 2$ orders less than the Comptonisation fluxes. The \textit{AstroSat} observation from 2018 is not able to resolve this emission line properly. However we were able to constraint its strength and it agrees well with the observations from the other epochs, within the $90\%$ confidence interval. In accordance with the low values of the emission line flux, major contribution to the total flux comes from the Comptonisation component. This is seen in Figure \ref{fig:fluxvar} where the individual spectral components are plotted, showing their variations during the observation period. We find that the model components closely follow the pattern of the total flux, rising, falling and then returning back to the original state over the three year time scale. All the flux values are tabulated in Table \ref{tab:flux}.

Akin to the strength of the emission line, we were unable to constrain the Gaussian width from the \textit{AstroSat} spectrum. However, analysis  with \textit{NuSTAR} and \textit{XMM-Newton} reveal a sudden change in the line width. \textit{NuSTAR} data from 2019 and August 2020 exhibit a broad iron line of width in the range $0.2$ - $0.3$ keV. But in December 2020, \textit{XMM-Newton} spectrum disclosed a swift reduction in its value, where we constrain it with an upper bound of $0.14$ keV. 

\begin{table}
    \centering
    \caption{Broadband X-ray flux and luminosity of Mrk 279, calculated in the energy range $0.001$ - $100.0$ keV, over the years from 2018 to 2020.}
    \label{tab:lum}
    \begin{tabular}{c c c c}
        \hline\hline
        Date of & Mission & Flux & Luminosity\\
        Observation & & ($\times 10^{-11}$ erg $cm^{-2}$ $s^{-1}$) & ($\times 10^{44}$ erg $s^{-1}$)\\
        \hline
        06-Feb-2018 & \textit{AstroSat} & $16.61_{\pm0.54}$ & $3.06_{\pm0.10}$ \\
        29-Oct-2019 & \textit{NuSTAR} & $25.00_{\pm2.64}$ & $4.61_{\pm0.49}$ \\
        03-Aug-2020 & \textit{NuSTAR} & $13.19_{\pm1.34}$ & $2.43_{\pm0.25}$ \\
        05-Aug-2020 & \textit{NuSTAR} & $12.05_{\pm0.92}$ & $2.22_{\pm0.17}$ \\
        11-Aug-2020 & \textit{NuSTAR} & $16.44_{\pm2.36}$ & $3.03_{\pm0.43}$ \\
        20-Dec-2020 & \textit{XMM-Newton} & $19.75_{\pm6.45}$ & $3.64_{\pm1.19}$ \\
        \hline 
    \end{tabular}
\end{table}

\begin{table}
    \centering
    \caption{The net continuum flux (NCM), warm corona flux (WCM) and the emission line (Gaussian) flux for Mrk 279 through the three year time scale. All values are in units of erg $cm^{-2}$ $s^{-1}$.}
    \label{tab:flux}
    \begin{tabular}{c c c c c}
        \hline\hline
        Date of & Mission & NCM & WCM & Gaussian\\
        Observation & & ($\times 10^{-11}$) & ($\times 10^{-11}$) & ($\times 10^{-13}$) \\
        \hline
        06-Feb-2018 & \textit{AstroSat} & $16.61_{\pm0.54}$ & $8.87_{\pm0.99}$ & $<4.26$ \\
        29-Oct-2019 & \textit{NuSTAR} & $24.96_{\pm2.65}$ & $13.57_{\pm2.10}$ & $5.37_{\pm1.43}$ \\
        03-Aug-2020 & \textit{NuSTAR} & $12.85_{\pm1.01}$ & $9.42_{\pm1.01}$ & $1.86_{\pm0.45}$ \\
        05-Aug-2020 & \textit{NuSTAR} & $12.02_{\pm1.06}$ & $9.18_{\pm0.63}$ & $2.23_{\pm0.33}$ \\
        11-Aug-2020 & \textit{NuSTAR} & $16.42_{\pm2.36}$ & $12.25_{\pm2.35}$ & $1.67_{\pm1.00}$ \\
        20-Dec-2020 & \textit{XMM-Newton} & $20.06_{\pm6.06}$ & $11.27_{\pm6.92}$ & $1.71_{\pm5.31}$ \\
        \hline 
    \end{tabular}
\end{table}

%%%%%%%%%%%%%%%%%%%%%%%%%%%%%%%%%%%%%%%%%%%%%%%%%%
\section{Summary and Discussion}
\label{sect:res}
\begin{figure}
    \centering
    \includegraphics[width=0.6\linewidth]{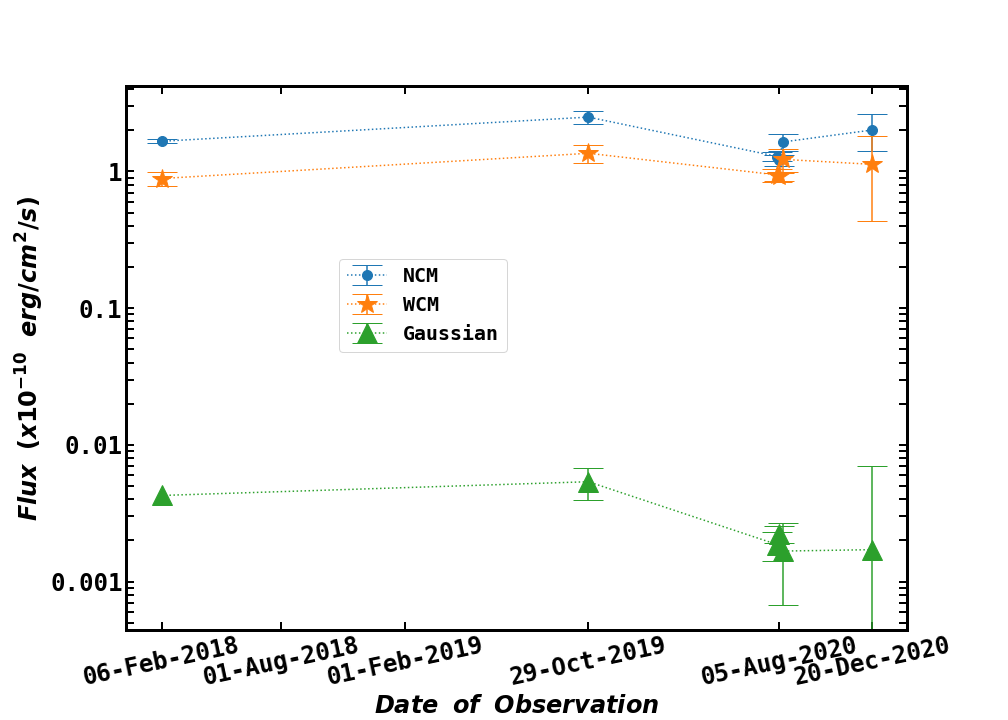}
    \caption{X-ray flux variations of Mrk 279 during the period from 2018-2020. The net continuum flux (NCM), warm corona flux (WCM) and emission line flux (Gaussian) are plotted in colors blue, orange and green, respectively.}
    \label{fig:fluxvar}
\end{figure}

\begin{figure}
    \centering
    \includegraphics[width=0.6\linewidth]{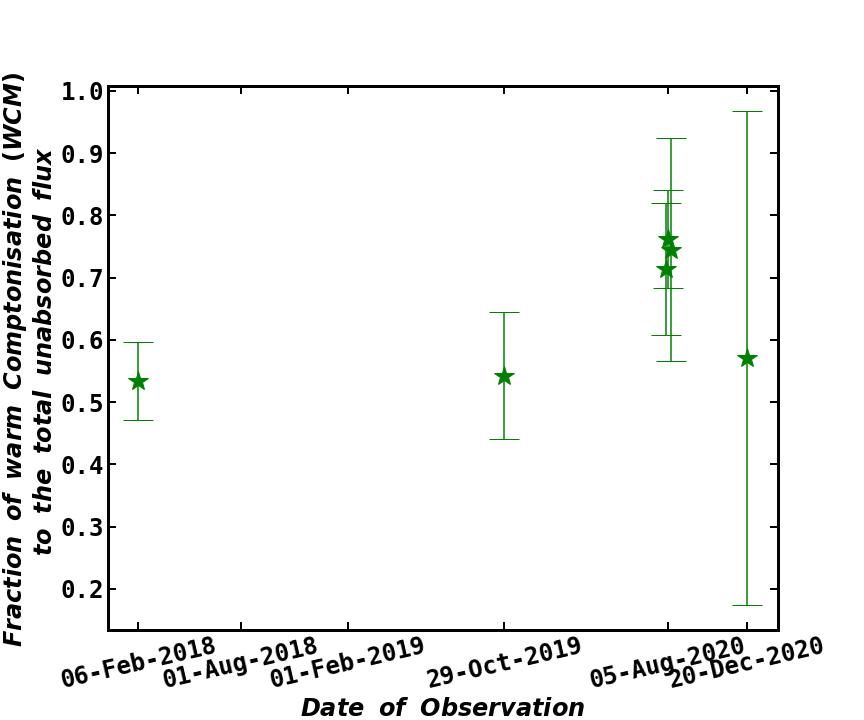}
    \caption{The fraction of warm Comptonised flux (WCM) to the total flux, showing its variations over the time period from 2018-2020.}
    \label{fig:ratiovsdate}
\end{figure}

\begin{figure}
    \centering
    \includegraphics[width=0.6\linewidth]{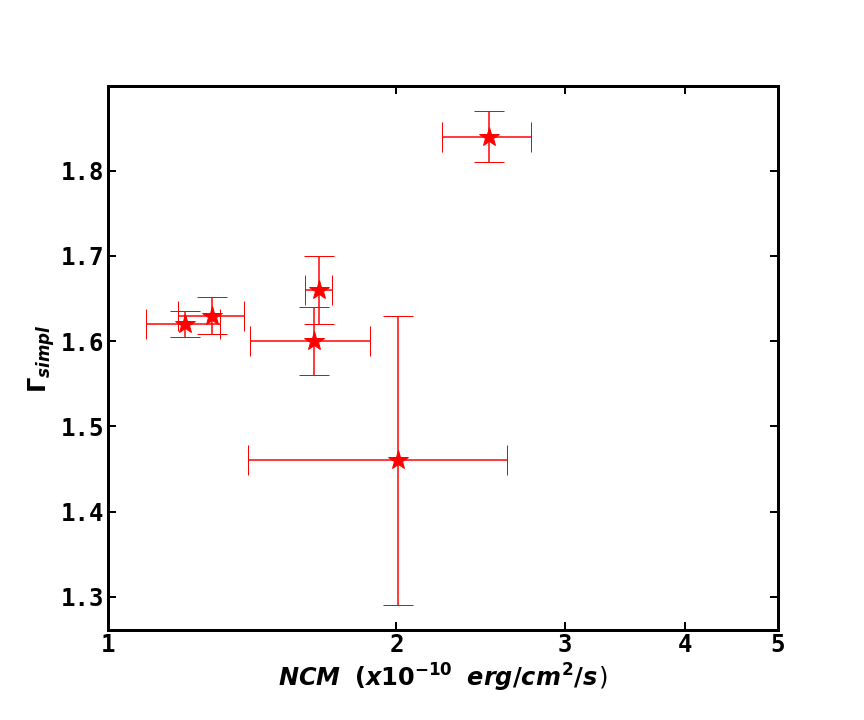}
    \caption{Hot photon index ($\Gamma_{simpl}$) plotted against the net continuum flux (NCM). The data do not seem to exhibit any significant correlation between the parameters; the Pearson correlation coefficient obtained was 0.41 with a two-sided p-value of 0.42.}
    \label{fig:gammavssimpl}
\end{figure}

In this work, we look at the X-ray spectrum of Mrk 279 over a period of three years to study the variations in its flux state. We also study the fluxes from the different spectral components in the spectrum and analyse their long term variations. We find that the flux values of all the spectral components of the source change considerably during this period, as is evident from Figure \ref{fig:fluxvar}, while in Figure \ref{fig:ratiovsdate} we plot the fraction of the warm Comptonised flux (WCM) to the total flux.

Unlike the spectral flux, the shape of the spectrum doesn't exhibit short term variations (the photon index remains at $\sim1.6$ for all \textit{NuSTAR} observations from August 2020). Over longer periods of time, the spectral photon index seem to follow the trend of the continuum flux at all epochs, except in December 2020 where the flux regains its previous high state but the index drops down further. 

Correlation between the continuum flux and photon index has been the subject of several previous studies \cite[E.g.,][]{singh1991observations, yaqoob1991x, dewangan200210, grupe2012remarkable, barua2020nustar}. Such a correlation could arise if the X-ray variability is due to the change in the seed photon population; in this case the physical property of the corona may vary with the flux. Another possibility in the two component scenario is of a super position of soft, variable power law associated with coronal emission and another harder spectral component with much less variability \citep[See:][]{shih2002continuum, fabian2003iron, markowitz2003long}. Nevertheless we found no significant correlation between the parameters; we obtained a Pearson correlation coefficient of 0.41 between the photon index ($\Gamma_{simpl}$) and the net continuum flux, with a two sided p-value of 0.42. Figure \ref{fig:gammavssimpl} shows the plot of $\Gamma_{simpl}$ versus the flux.

In conclusion we find that, in agreement with the previous studies, Mrk 279 shows significant variations in its flux state and hence, its luminosity. This pattern is closely followed by the spectral components as well. We also notice that the photon index and hence the spectral shape follow the flux variations over longer periods of time. However, neither the index or the spectral shape is seen to exhibit the short term changes seen in the X-ray flux. Subsequently we plan to include UV data from \textit{AstroSat}'s UVIT mission to obtain a broadband spectral model and analyse the correlations between variabilities in the different energy bands.

\normalem
\begin{acknowledgements}
We thank the anonymous referee for the valuable suggestions and comments that improved the manuscript. Authors KA and KJ acknowledge the financial support from ISRO (Sanction Order:No.DS\_2B-13013(2)/11/2020-Sec.2). KA thank the IUCAA visiting program. KJ acknowledges the associateship program of IUCAA, Pune. KA is thankful to colleagues Suchismito Chattopadhyay, Prajjwal Majumder and Sree Bhattacherjee for the useful discussions during the course of the work. This publication uses the data from the AstroSat mission of the Indian Space Research Organisation (ISRO), archived at the Indian Space Science Data Centre (ISSDC) and may be accessed using the Observation Id 9000001886. This research has made use of data from the NuSTAR mission, a project led by the California Institute of Technology, managed by the Jet Propulsion Laboratory, and funded by the National Aeronautics and Space Administration. Data analysis was performed using the NuSTAR Data Analysis Software (NuSTARDAS), jointly developed by the ASI Science Data Center (SSDC, Italy) and the California Institute of Technology (USA). This research has made use of \textit{XMM-Newton} data software provided by the High Energy Astrophysics Science Archive Research Center (HEASARC), which is a service of the Astrophysics Science Division at NASA/GSFC.
\end{acknowledgements}
  
\bibliographystyle{raa}
\bibliography{main}

\end{document}